# Metastable low energy states in $TiS_2$, $TiSe_2$, $TiTe_2$ systems predicted with evolutionary algorithms

Oleg D. Feya[1,2], Dmitri V. Efremov[1]

1. *Leibniz Institute for Solid State and Materials Research, IFW Dresden, Helmholtzstraße 20, 01069 Dresden, Germany*
2. *Kyiv Academic University, 03142, Kyiv, Ukraine*

**Abstract.** *We report a systematic study of low energy metastable states in van der Waals semimetals $TiS_2$, $TiSe_2$ and $TiTe_2$ within the DFT theory by means the evolutionary search algorithm. We find a big difference between $TiSe_2$, $TiS_2$ and $TiTe_2$ in low energy metastable states. While several metastable states exist in $TiSe_2$ and $TiS_2$, no low energy metastable states were found in $TiTe_2$. We show that some of the obtained metastable states can be identified as charge density wave (CDW). We argue that existence of the low energy metastable phases indicates that emergence CDW as a ground state in these compounds.*

## Introduction

Transition metal dichalcogenides (TMDs) are a class of layered quasi-two-dimensional van-der-Waals materials described the general formula $MX_2$, where M is a transition metal (Ti, Mo, V, Nb, etc.), and X is a chalcogen (Te, S, Se, etc.) [1–3]. The consistently high interest in these materials has been fueled by the discovery of graphene [4] and the recent discovery of the topological nature of the electronic states in some TMDs [5]. TMDs consist of stacked trilayer X-M-X , and have hexagonal or trigonal symmetry. These trilayers are held together by the weak van der Waals forces, which allows exfoliation of the individual trilayers and the deposition of these layers onto various substrates [6]. This simple method of the obtaining of nanostructures together with rich physical properties of the TMD make them promising materials for spintronics, nanoelectronics, renewable power generation, biochemical applications [3], as well for valleytronics – brand new approach for quantum computations [7–9].

A key feature common to many TMD materials is the charge density wave (CDW) ordering, which often occurs in the vicinity of superconductivity [10]. A typical example is $1T-TiSe_2$ in which the transition to the CDW state takes place at $T_{CDW} \sim 200$ K [11]. Below $T_{CDW}$ $TiSe_2$ crystallizes into a 2x2x2 CDW superlattice form with the P-3c1 (#165 space symmetry) symmetry, while undistorted crystal at $T>T_{CDW}$ has the P-3m1 (#164) symmetry. Below the transition the resistivity strongly increases [10].

Related compounds $1T-TiS_2$ and $1T-TiTe_2$ are obtained from $TiSe_2$ by isovalent substitution of Te and S for Se, respectively. The compounds have the same crystal structure as $1T-TiSe_2$ at ambient conditions. However, unlike $TiSe_2$, $1T-TiTe_2$ do not exhibit any CDW in bulk at low temperatures [12,13]. Recently, it was shown that (2 × 2) CDW can emerge only in monolayers of $TiTe_2$ (P.Chen et al). The experimental situation with $TiS_2$ is more controversial. It is still debated whether it has semiconductor or semimetallic nature [14]. A part of the phase diagram $TiSe_{2-x}S_x$ up to x<0.34 was recently studied in [15] by means of ARPES and STM. The authors have seen that CDW is gradually suppressed. Extrapolating their results, the authors concluded that CDW vanishes under doping close to x = 1.

As charge density modulation and structure distortion in the CDW state happen together, the driving force of the CDW transition is still debated. A few mechanisms were proposed, including phonon condensation [16], excitonic mechanism [17], band Jahn-Teller instability [18], orbital ordering [19], interplay of cooper and particle-hole instabilities [20].

The aim of this work is to study of the tendency towards CDW in the row Ti(S, Se, Te)$_2$ by analyzing of the ground state and low-energy metastable states corresponding to the low-energy local minima of the Kohn-Sham functional. To obtain the local minima, the evolutionary search method with fixed composition is used in above mentioned systems [21]. This method was previously successfully used to predict new phases of layered materials like VSe$_2$ [22], PS$_2$ [23], FeS$_2$ [24], TiTe$_2$ under pressure [25], and to treat 2D-systems such as borophene [26], phagraphene [27], 3d transition metal monocarbides [28]. For the obtained structures, we investigate the effect of the van der Waals correction to the Kohn-Sham functional on the parameters of predicted structures.

**Computational Methodology**

We use USPEX code to predict crystal structures of the systems with mixed chalcogen composition and building of stability phase diagrams [29,30]. Calculations of the low-energy metastable configurations for the chosen composition is implemented in this code. We used the variable search mode with minimum of 12 atoms and maximum of 18 atoms for the 1$^{st}$ generation. It was produced by a random structure generator and consisted of 200 structures. Each of the next generations had 100 compositions, which were obtained by applying 40% heredity, 20% softmutation, 10% lattice mutation and 30% random structure generator. Each structure was carefully relaxed in four stages, starting with low precision.

For structure relaxations in USPEX we used the density functional theory (DFT) [31,32] implemented in the VASP code [33–35]. The Perdew-Burke-Ernzerhof (PBE) generalized gradient approximation (GGA) [36] and the projector-augmented wave (PAW) [37,38] in DFT were also employed for these calculations. We set the kinetic energy cutoff to 350 eV. Uniform Γ-centered k-meshes with reciprocal-space resolution of $2\pi \times 0.07$ Å$^{-1}$ were used for Brillouin zone sampling. In the post-processing we carefully relaxed all low-energy structures in USPEX output with 6x6x3 k-points mesh and energy cutoff 400 eV. Band structure, density of states and fermi-surface calculations we performed with FPLO package [39,40] within GGA approximation, 12x12x6 mesh and 100 points between high-symmetry points in k-space. To build fermi surfaces we had expanded the k-grid to 40x40x20 points.

**Results and discussions.**

As the first step, we compare the theoretically obtained 1T structures having P-3m1 symmetry (CdI$_2$ phase) and the space group #164, with the correspondent experimental ones. We use the structures relaxed with PBE functional, PBE-D3(BJ) with the Becke-Jonson dumping [41,42] and the Grimme zero dumping method PBE-D3(ZD) [42]. We also perform calculations with dispersion correction to treat the van der Waals forces. All considered functionals result in a semimetalic state with shallow electron and hole bands for all considered compounds. It matches with the experiments for TiSe$_2$ and TiTe$_2$. However, some experiments in TiS$_2$ indicate a semiconducting state with a small gap in contrast to the semimetallic ground state within the

DFT theory [43].

The results are collected in Table 1. Using of PBE functional for the structure relaxation leads to overestimation of the crystal structure parameter c by ~10% for $TiS_2$ and $TiSe_2$, and by 3.3% for $TiTe_2$. The Becke-Jonson correction slightly underestimates the parameter c with applying too huge interlayer connection. The zero dumping dispersion correction method gives that best matching with experimental data for all these materials. Similar results were reported for the PBE functional for numerous TMD structures [44] and for the hybrid PBE0 functional for Ti, Hf and Zr disulphides [45].

| Parameter | $TiS_2$ | $TiSe_2$ | $TiTe_2$ |
|---|---|---|---|
| Experimental data | | | |
| a | 3.407 | 3.540 | 3.768 |
| c | 5.695 | 6.010 | 6.460 |
| PBE | | | |
| a | 3.409 (~0%) | 3.534 (-0.17%) | 3.799 (+0.79%) |
| c | 6.487 (+14.08%) | 6.736 (+12.08%) | 6.628 (+2.60%) |
| PBE-D3(BJ) | | | |
| a | 3.368 (-1.14%) | 3.479 (-1.72%) | 3.720 (-1.24%) |
| c | 5.542 (-2.68%) | 5.895 (-1.91%) | 6.316 (-2.29%) |
| PBE-D3(ZD) | | | |
| a | 3.400 (-0.21%) | 3.520 (-0.56%) | 3.772 (+0.11%) |
| c | 5.687 (-0.14%) | 6.065 (+0.92%) | 6.421 (-0.60%) |

**Table 1**. Structural parameters (in Angstrom) for pure $TiX_2$ systems obtained with PBE functional with and without disperse correction, and difference with experimental data

Now we investigate the energy variation in the experimentally observed 3q-CDW. We use three exchange-correlation approximations: LDA, GGA and GGA-D3 without spin-orbit coupling. The spin-orbit coupling gives small contribution to the total energy [46] and has high computational costs for the evolutionary calculations. We plot the energy per unit cell as the function of the atom displacement in the way undistorted #164 structure – 3q-CDW. For this purpose, we use undistorted 2x2x2 supercells with parameters of the experimental structure [11,47], and fully relaxed structure with the corresponding functionals. For $TiSe_2$ it is means a=b=7.080 Å and c=12.020 Å (doubled experimental values from table 1), $\delta_{Ti}$=0.012a=0.085Å, and $\delta_{Ti}/\delta_{Se}$ =3. Here $\delta_{Ti}$ and $\delta_{Se}$ are displacements of Ti and Se atoms relative to #164 structure. The model of such structures is shown on Fig 1. The parametrization is adopted from [47]. The same model is used for construction of 3Q-CDW for $TiS_2$ and $TiTe_2$.

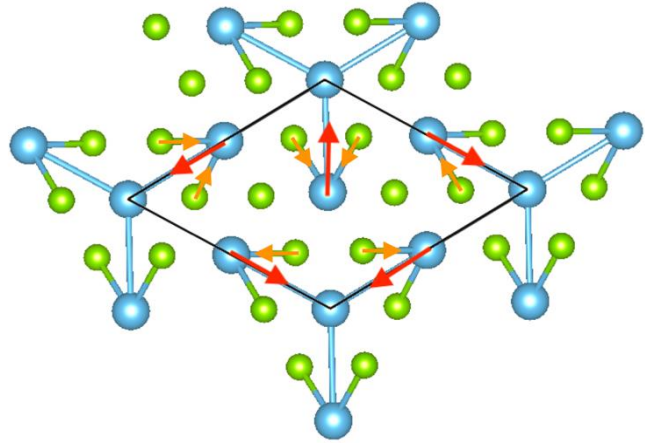

**Figure 1.** Model of TiSe$_2$ - CDW-structure. Blue circles indicate Ti atoms, green – Se atoms. Solid lines indicate unit cell, which is 2x2 of original undistorted unit cell. The red arrows indicate directions of the Ti atoms displacements. The orange arrows indicate directions of the chalcogen atoms displacements.

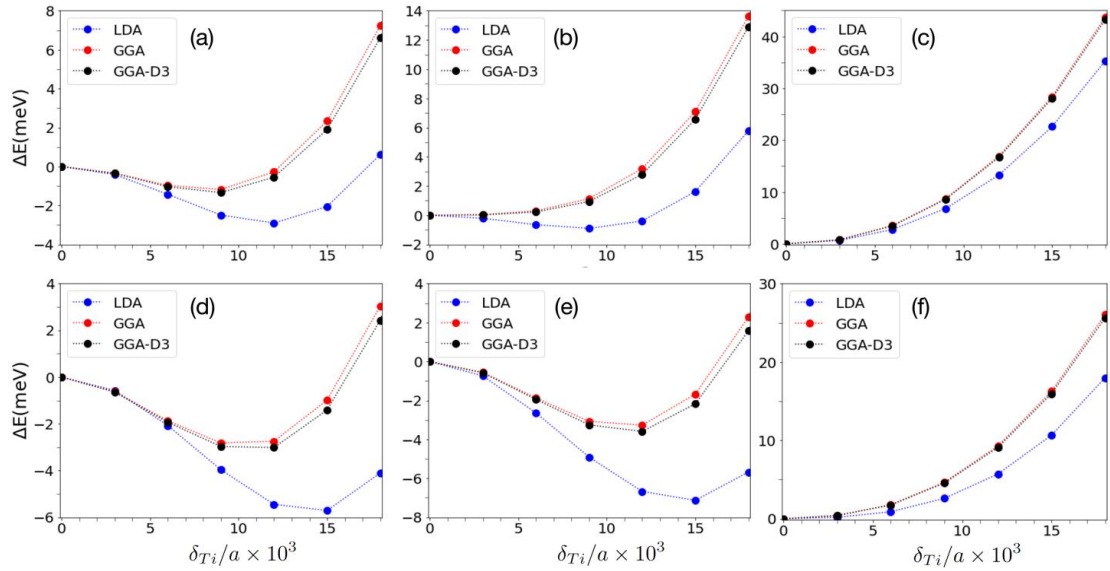

**Figure 2.** Energy variation per formula unit in meV as a function of applied distortions to experimental (a) TiS$_2$; (b) TiSe$_2$; (c) TiTe$_2$ and theoretical (d) TiS$_2$; (e) TiSe$_2$; (f) TiTe$_2$ with 164 symmetry. Lattice constants and atoms positions are frozen. The displacements of Ti $\delta_{Ti}$ is expressed in units of the structure constant. The value of the chalcogen displacement is approximately three times smaller.

In Fig 2 (a-c) we present the energy differences for TiS$_2$, TiSe$_2$ and TiTe$_2$ with fixed experimental lattices constants and frozen atom positions. On each step we move Ti and chalcogen atoms from original positions in #164 structure in directions of the experimental CDW displacements (Fig. 1) Within these functional we see that the minimum corresponds to the CDW state for TiS$_2$, TiSe$_2$ and undistorted state for TiTe$_2$. The obtained results experimental observations of CDW displacements, but GGA, GGA-D3 and LDA functionals are able to "catch" such distortion (Fig2 a) at $\delta_{Ti}$ ~ 0.061 Å for GGA and $\delta_{Ti}$= 0.082 Å for LDA. For TiSe$_2$ GGA and GGA-D3 functionals show the minimum corresponds to the undistorted lattice. In contrast, LDA demonstrates the minimum at for the displacement $\delta_{Ti}$ ~ 0.064 Å, which is less $\delta_{Ti,exp}$ ~ 0.080 Å (Fig 2b). For TiTe$_2$, all functionals have the minimum at $\delta_{Ti} = 0$ (Fig 2c). It agrees with experimental observation of the absence of CDW in bulk TiTe$_2$.

As a next step we investigate using the same method the theoretical lattices with the fixed lattice constants given in Table 1. All functionals successfully predict the minimum of the functional for CDW for TiSe$_2$ (Fig 2e) and for TiS$_2$ (Fig 2d), but no CDW for TiTe$_2$ (Fig 2f). Instabilities in TiSe$_2$ for GGAs occurs at $\delta_{Ti}= 0.085$ Å which is close to experimental CDW, but LDA shifts the distortion to $\delta_{Ti}$~ 0.106 Å with energy gain 7.13 meV/f.u compared to 3.25 meV/f.u and 3.59 meV/f.u for GGA and GGA-D3. Again, it agrees with the experimental observation of emergence of CDW in TiSe$_2$, but not in TiTe$_2$.

The full relaxation of the atoms with the fixed experimental lattice constants leads to distortions $\delta_{Ti}= 0.080$ Å for GGA ($\delta_{Ti}/\delta_{Se} = 3.35$), 0.076 Å (2.76) for LDA, 0.088 Å (3.32) for GGA-D3. As we see, in this case all functionals give close to experimental $\delta_{Ti}$ value, but GGAs overestimate ratio between Ti and Se displacements, while LDA underestimate it. The picture changes for theoretical lattice constants. With LDA any distortions disappear, as well as for zero dumping vdW corrections. GGA has Ti displacement 0.073 Å and the ratio equals 3.37.

The description of CDW distortions in TiSe$_2$ within various exchange-correlation functionals was previously discussed in [46–48]. It has been shown that the local density approximation (LDA) doesn't "catch" this instability [49], as well as GGA+U [46]. This happens due to the strong contribution of the long-range interactions. Our results for TiSe$_2$ agree with results of these works. Our further investigation has shown similar results for TiS2, while for TiTe$_2$ for all functionals only the undisturbed state is stable.

| Space group symmetry | Lattice constants (Å) | Energy difference (mEv/f.u) | Ti-Se bonds (Å) | Ti-Ti distances (Å) | $\delta_{Ti}$ (Å) |
|---|---|---|---|---|---|
| TiS$_2$ | | | | | |
| 13 | a = 5.925<br>b = 3.408<br>c = 6.786 | -25.6 | 2.378<br>2.430<br>2.488 | 3.358<br>3.408<br>3.481 | 0.062 |
| | a = 10.243<br>b = 10.243<br>c = 19.045 | -24.2 | 2.372<br>2.404<br>2.467 | 3.362<br>3.414<br>3.467 | 0.052 |
| 165 | a = 6.828<br>b = 6.828<br>c = 12.962 | -16.3 | 2.344<br>2.429<br>2.539 | 3.310<br>3.419<br>3.518 | 0.104 |
| 164 | a = 3.409<br>b = 3.409<br>c = 6.487 | 0 | 2.430 | 3.409 | 0 |
| TiSe$_2$ | | | | | |
| 13 | a = 6.090<br>b = 3.539<br>c = 6.700 | -16.7 | 2.540<br>2.559<br>2.598 | 3.491<br>3.539<br>3.554 | 0.032 |
| 12 | a = 6.093<br>b = 3.541<br>c = 6.675 | -16 | 2.563<br>2.568 | 3.519<br>3.530 | - |
| 2 | a = 6.115<br>b = 7.244<br>c = 3.534 | -14 | 2.515<br>2.563<br>2.622 | 3.474<br>3.534<br>3.594 | - |
| 165 | a = 7.081<br>b = 7.081<br>c = 13.464 | -11.5 | 2.497<br>2.565<br>2.647 | 3.457<br>3.543<br>3.624 | 0.083 |
| 2 | a = 7.075<br>b = 7.059<br>c = 7.595 | -4.2 | 2.517<br>2.561<br>2.614 | 3.477<br>3.524<br>3.535<br>3.601 | - |
| 164 | a = 3.534<br>b = 3.534<br>c = 6.736 | 0 | 2.566 | 3.534 | 0 |
| TiTe$_2$ | | | | | |
| 164 | a = 3.799<br>b = 3.799<br>c = 6.628 | 0 | 2.776 | 3.799 | 0 |
| 166 | a = 3.776<br>b = 3.776<br>c = 22.107 | +43 | 2.784 | 3.776 | 0 |

Table 2. The ground and metastable states of TiX$_2$ system, obtained by USPEX package with GGA PBE

| Structure | Space group symmetry | Lattice constants (Å) | Energy difference (mEv/f.u) | Ti-Se bonds (Å) | Ti-Ti distances (Å) |
|---|---|---|---|---|---|
| TiSe$_2$ | 15 | a = 5.936<br>b = 3.439<br>c = 13.075 | -6.8 | 2.491<br>2.512<br>2.540 | 3.405<br>3.439<br>3.456 |
| TiSe$_2$ | 12 | a = 5.943<br>b = 3.430<br>c = 5.801 | -5.6 | 2.502<br>2.521 | 3.429<br>3.431 |
| TiSe$_2$ | 164 | a = 3.441<br>b = 3.441<br>c = 5.798 | 0 | 2.512 | 3.441 |
| TiSe$_2$ | 12 | a = 6.055<br>b = 3.522<br>c = 6.060 | -1.9 | 2.549<br>2.558 | 3.502<br>3.522 |
| TiSe$_2$ | 164 | a = 3.520<br>b = 3.520<br>c = 6.065 | 0 | 2.557 | 3.520 |

Table 3. Local energy minima for TiSe$_2$ system, obtained by USPEX package with LDA and GGA-D3

As the next step we investigate all possible low energy local minima of for TiS$_2$, TiSe$_2$ and

TiTe$_2$ using the evolutionary search followed by the VASP relaxation. We used only PBE potential to reduce computational cost. We present all low energy local minima of the DFT functionals, obtained in USPEX for TiS$_2$, TiSe$_2$ and TiTe$_2$ in Table 2. They correspond to the ground state and to the metastable states.

First, let us discuss the results for TiSe$_2$. All energies are given with respect to the state with the space group symmetry #164 corresponding to the experimental high temperature structure. and the state with the space group symmetry #165 corresponding to CDW are only metastable states within these calculations. The states #13 and #12, which have lower energy than #164, can be obtained from #164 by distortion of Ti atoms. The state #13 can be interpreted as a single-q CDW. The structure in the real space are demonstrated in Fig. 3. The propagation vector of the distortion corresponds to the nesting vector of connecting the hole and the one of the electron pockets in the Brillioune zone (see Fig 4). An example of a such structure is #13-TiSe$_2$ and #13-TiS$_2$ (Fig3b as the model structure for both). They can be obtained from the undistorted structure #164 by shifting the Ti atoms along the short diagonal of the unit cell in one direction, and atoms, connected by the long diagonal, in the opposite direction. With this rearrangement the symmetry changes to monoclinic P2/c (#13) from trigonal P-3m1 (#164). The lattice constant is doubled in *a* direction. For TiSe$_2$ length of distortion is 0.032 Å, while in TiS$_2$ the Ti atoms moving by 0.062 Å. Similar single-q CDW interpretation admits #2 structure (Fig. 3c).
Besides the experimentally found in TiSe$_2$ 3q-CDW with the space group #165 we have found a possibility of realization of another type of 3q-CDW with the space group #148 in TiS$_2$.

Now we check how low energy metastable states distribution changes with the exchange functionals. We perform series of fixed-composition calculations for LDA, PBE and PBE-D3 for TiSe$_2$ system. Results on LDA and PBE-D3 pseudopotentials we show in Table 3. Structure #15-TiSe$_2$, obtained by LDA, has generally the same pattern, as previously discussed #13, but its unit cell consists of 2 shifted layers. Another difference in vdW gap, which is ~2.5% lower than for #13 structure obtained by PBE pseudopotential. Size of this gap defines the main difference between #12 structures calculated with all the potentials. Vector constant c changes from 5.801 Å for LDA, and grows to 6.060 Å for PBE-D3 with the biggest of 6.700 Å for PBE. Which is expected due to data collected in Table 1. Energy difference between #12 and #164 for PBE-D3 is the lowest for all discussed functionals. Also, we have included there CDW triple-q structure with #165 symmetry to compare energy differences. One can see the PBE pseudopotential for TiSe$_2$ results in plenty local minima, while PBE plus dispersion correction and LDA fails to find some of the structures as well as "catch" CDW-3q distortion. Typical low-energy local minima are shown on Fig 3. They show same pattern for two considered systems, while TiTe$_2$ stays separately having no local minima lower in energy than #164 undistorted structure.

The change of the structures leads to the Fermi surface restructuring. GGA potential treats TiS$_2$ as semi-metal with three small electron pockets in L high-symmetry points of Brillouin zone and hole-pocket in Γ point (Fig4 a, c). All mentioned above distortions can be described as single-q changes of Brillouin zone, which leads to shift of these pockets on the length of Q-vector. One electron pocket matches with hole-pocket in Γ point, while two others superimpose, forming tiny "propeller" pocket (Fig4 b). Fermi surface almost disappears (not showing on the figures), and density of states drops near Fermi-level (Fig5 a,b). Same process occurs in TiSe$_2$ zig-zag structure. As result of Fermi-surface restructuring, one L-pocket superimpose Γ-hole-pocket, and another electron pockets form one big propeller pocket (Fig 4

e,h). Central cylinder of Fermi-surface evolves from "dumbbell" form with thickened region at the center of Brillouin zone to cylinder with smoothly changing diameter, and only one "propeller" pocket remains. This process accompanied with lowering density of states near Fermi level compared with #164 $TiSe_2$ (Fig 5c,d).

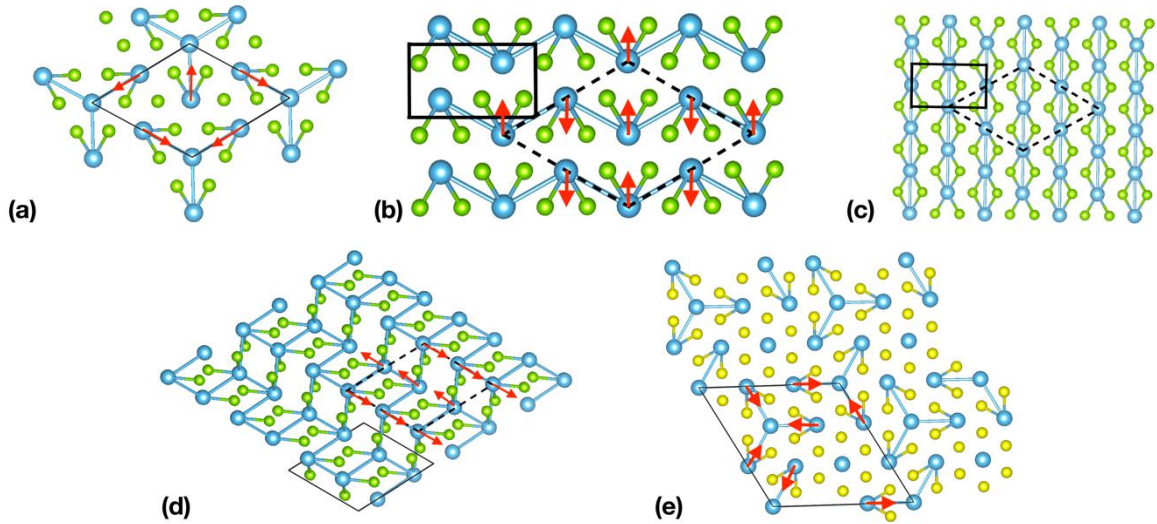

**Figure 3.** Top view of the low energy metastable states with PBE functional. (a) Classical $TiSe_2$ CDW structure (space symmetry #165); (b) Single-q $TiSe_2$ structure (space symmetry #13); (c) distorted $TiSe_2$ structure (#12); (d) Distorted $TiSe_2$ (#2); (e) 3x3 $TiS_2$ (#148). Blue circles indicate Ti atoms, green – Se atoms, yellow – S atoms. Solid lines show for unit cell, dashed lines show the 2x2 pattern. The red arrows indicate directions of the atom displacements. The shortest atoms bonds are shown by blue, green and yellow lines

Structure with #12 symmetry occurs in calculations for $TiSe_2$ with all used functionals (LDA, PBE and PBE-D3). Its lattice vectors *a* and *b* align the diagonals of original #164 unit cell. *b* vector corresponds shorter diagonal, and its length decreases by 0.021 Å compared to this diagonal (Fig 3c). Density of states almost coincide with those of #164 structure (Fig 5c,d). Space group #12 is a subgroup of #164 symmetry, so we consider this structure as variation of original $TiSe_2$ with no sign of charge density wave distortion.

Structure with symmetry #2 (Fig 3d) is more complex case than zig-zag #13 and #12 structures. Generally, it shows the same zig-zag pattern as #13 structure with 3.483 Å Ti-Ti bonds length (compared to 3.491 Å for #13-$TiSe_2$ and 3.534 Å for #164-$TiSe_2$), but the distance between these Ti-waves now not identical. While for #13 -$TiSe_2$ this distance equals 3.539 Å, for #2-$TiSe_2$ there are two types of bonds – 3.524 Å and 3.535 Å. This structural changes accompanied by dramatical modification of Fermi surface (Fig 4f,i). Now it is composed of central electron ellipsoid-like region surrounded by four electron pockets. Hole conductance regions are gathered on the edges of Brillouin zone. As it is impossible to get such pattern from q-vectors restructuring of #164-$TiSe_2$ Fermi surface, we consider #2-$TiSe_2$ as stand-alone modification of original $TiSe_2$.

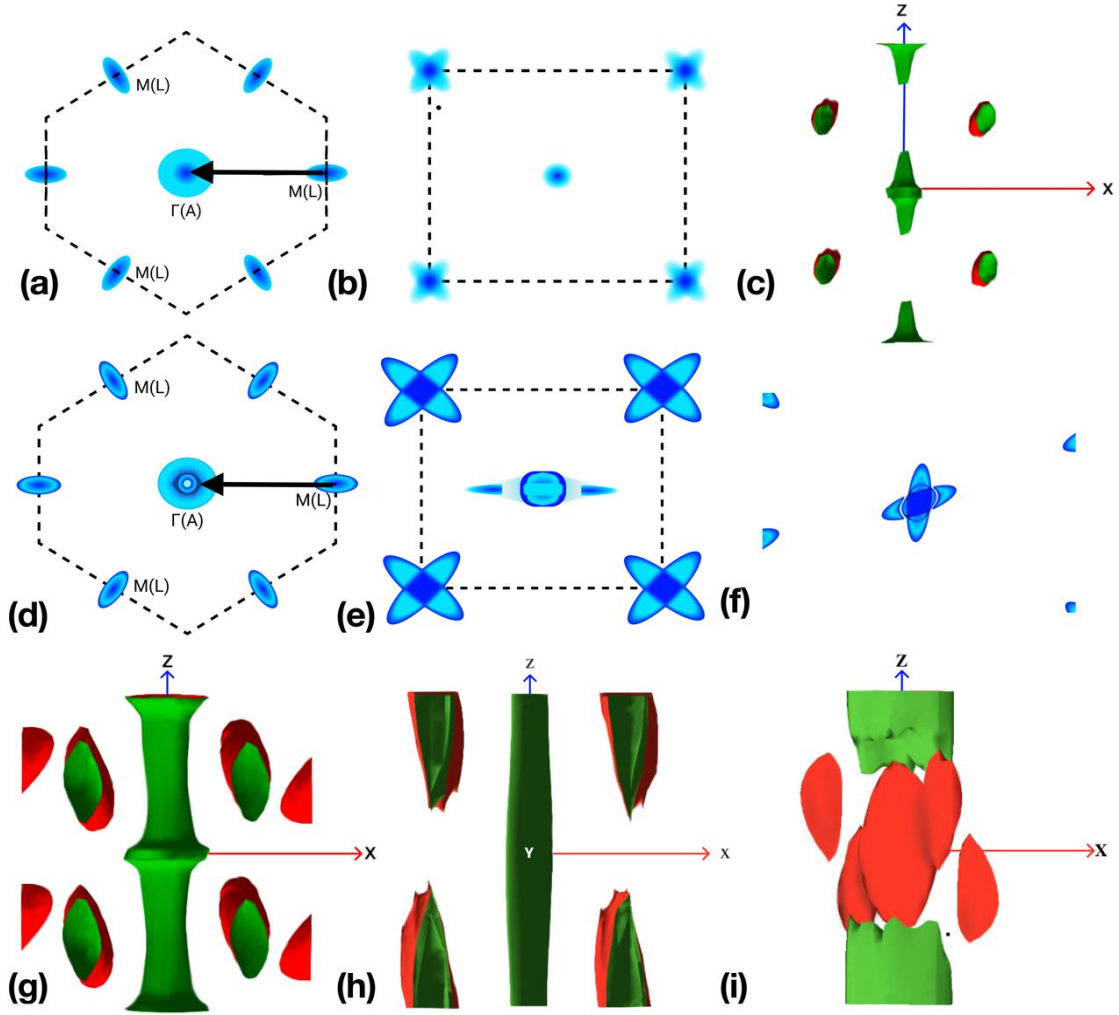

**Figure 4**. (a, b) 2D cut of Fermi-surface for $k_z=0$ for undistorted $TiS_2$ (#164) and zig-zag structure (#13); (c) Fermi-surface of $TiS_2$; (d, e) 2D cut of Fermi-surface normal to Z direction for undistorted $TiSe_2$ (#164) and zig-zag structure (#13); (g,h) Fermi-surface of $TiSe_2$ and zig-zag structure; (f-i) 2D cut and 3D Fermi surface of #2 distorted structure. Black dashed line indicates Brillouin zone boundary, and black arrows – nesting vectors.

Structure #148-$TiS_2$ (Fig 3e) characterized by triple-q pattern, which is corresponds to the same of classic CDW $TiSe_2$ distortion (Fig 3a). It has huge supercell 3x3x3 while triple-q $TiSe_2$ with 2x2x2 supercell is much smaller. Ti-S bonds inside these triangles have 2.372 Å length, which is shorter compared to undistorted 2.430 Å Ti-S bonds in $TiS_2$. Ti-atoms are moved by 0.052 Å along *a* and *b* lattice vectors. Except these triangles, its unit cell contains stand-alone Ti-S hexagon with slightly distorted Ti-S bonds from original structure. Overall, this structure share similar patterns with previously predicted with machine learning techniques 2D 3x3 and 4x4 $TiS_2$ superlattices [50]. This structure is of particular interest – it has very low number of states at the Fermi level (Fig5b), which may indicate its CDW nature.

After isovalence substitution of Te atoms on positions of S and Se atoms in all considered structures and careful relaxation with PBE functional, all these distortions disappear and the structures transform into #164-$TiTe_2$. It may indicate the importance of spin-orbit coupling effect on existence of CDW patterns.

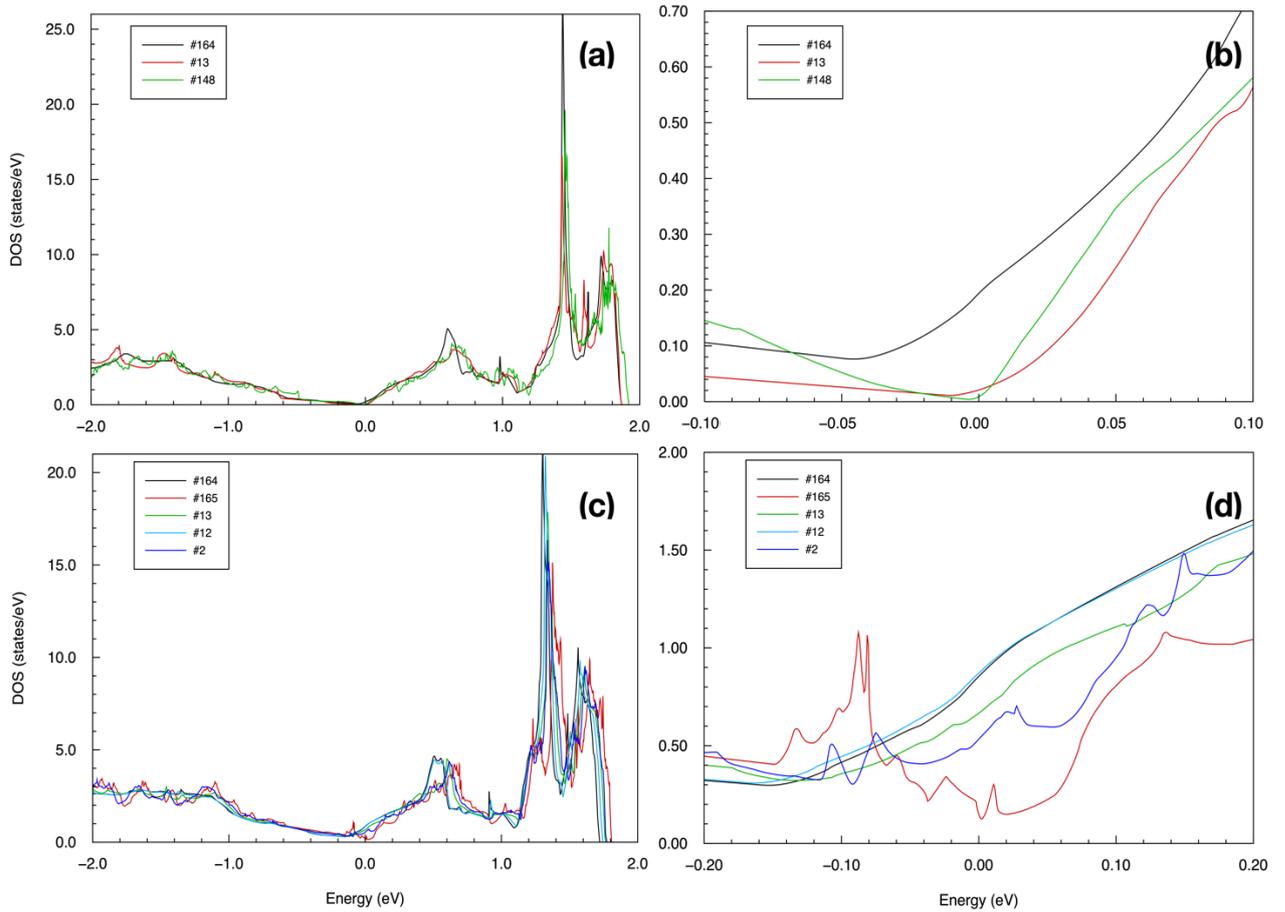

**Figure 5**. Density of states of typical local minima of $TiS_2$ and $TiSe_2$ systems: (a,b) undistorted #164, zig-zag #13 and #148 $TiS_2$ structures; (c,d) undistorted #164, triple-q #165, zig-zag #13, distorted #12 and #2 $TiSe_2$ structures. Shown ranges [-2; 2] eV and [-0.2; 0.2] eV to compare DOSs at Fermi-level.

**Conclusions**

Here we have shown that combination of evolutionary algorithms and DFT, applied on various dichalcogenides systems, can overall predict the existence of numerous metastable CDW phases competing in energy with well-known distortions. For bulk $TiS_2$, which has no experimental CDW phases and shows metal-semiconductor transition, we have predicted several metastable structures with space symmetry #13 and #148, which close in energy to original #164-structure. $TiSe_2$, according to our findings, has not only well-known #165 CDW distortion, but single-q phase with #13 space symmetry and various metastable phases with another symmetries (#2, #12). While $TiTe_2$, which has no CDW transition and strong spin-orbit coupling due to nature of Te, have no metastable phases around original structure.

Corrections to GGA functionals are important. Calculated parameters of dichalcogenide structures are in best correspondence with experiment for GGA-D3 than for other functionals. For evolutionary search it shows almost absence of metastable local minima, which are not observed in experiments.

**Acknowledgment.**

We thank Sergey Borisenko for illuminating discussions. Also we thank Ulrike Nitzsche for technical assistance and Katerina Guslenko for help with graphical design. This work was supported by BMBF through the project UKRATOP. We also acknowledge RSF-DFG project 405940956 for financial support.

# Bibliography.